%% file: main.tex
\documentclass[runningheads]{llncs}
\usepackage[utf8]{inputenc}
\usepackage{listings}
\usepackage{todonotes}
\usepackage{graphicx}
\usepackage{caption}
\usepackage{subcaption}

\title{PSY-TaLiRo: A Python Toolbox for Search-Based Test Generation for Cyber-Physical Systems}
\author{Quinn Thibeault
    \and Jacob Anderson
    \and Aniruddh Chandratre
    \and Giulia Pedrielli
    \and Georgios Fainekos}
\authorrunning{Quinn Thibeault et al.}
\titlerunning{PSY-TaLiRo}
\date{May 2021}
\institute{Arizona State University, Tempe AZ, 85281, USA}

\begin{document}
\maketitle

\begin{abstract}
    In this paper, we present the Python package PSY-TaLiRo which is a toolbox for temporal logic robustness guided falsification of Cyber-Physical Systems (CPS). 
    PSY-TaLiRo is a completely modular toolbox supporting multiple temporal logic offline monitors as well as optimization engines for test case generation.
    Among the benefits of PSY-TaLiRo is that it supports search-based test generation for many different types of systems under test.
    All PSY-TaLiRo modules can be fully modified by the users to support new optimization and robustness computation engines as well as any System under Test (SUT).
    
    \keywords{Falsification, Cyber-Physical Systems, Search-Based Test Generation}
\end{abstract}

\section{Introduction}
	
	Requirements falsification for Cyber-Physical Systems (CPS) has gained prominence in recent years as a practical way to test and debug industrial complexity models and systems \cite{TuncaliEtAl2018nfm,SankaranarayananEtAl2017sigbed,MenghiEtAl2020icse,YamaguchiKDS2016fmcad}.
	Since the automotive industry was an early adopter of the falsification technology \cite{KapinskiEtAl2016csm}, many of the benchmark CPS models driving the research were MATLAB/Simulink models \cite{ChutinanB02fordtech,StrathmannO15arch,JinEtAl2014hscc,HoxhaAF14arch2}.
	As a result, some of the academic falsification tools are MATLAB tools: Breach \cite{donze10cav}, S-TaLiRo \cite{AnnapureddyLFS11tacas}, and ARIsTEO \cite{MenghiEtAl2020icse}.
	Other academic falsification tools that participate in the ARCH falsification competition \cite{ErnstEtAl2020archComp} are {\sc FalStar} \cite{ZhangEtAl2018cadics} (Java/Scala), zlscheck \cite{zlscheck} (OCaml with Zelus models), and falsify \cite{AkazakiEtAl2018} (ChainerRL \cite{ChainerRL} Python Library for reinforcement learning calling MATLAB functions).
	
	However, as the autonomy and robotics research communities (and even industry) increasingly adopt Python as the preferred language for prototyping, there is a need for a falsification toolbox natively in Python.
	An all Python/C++ falsification framework would resolve any computational inefficiencies and compatibility issues of calling Python from MATLAB and/or vice versa.
	The PSY-TaLiRo (or $\Psi$-TaLiRo) toolbox, wihch stands for Python SYstems' TemporAl LogIc RObustness, addresses exactly this need.
	It is a fully modular and extensible toolbox for temporal logic guided falsification which mirrors the S-TaLiRo \cite{AnnapureddyLFS11tacas} structure.
	Namely, the users can easily call different temporal logic robustness computation engines (e.g., TLTk \cite{tltk}, RTAMT \cite{rtamt}), optimizers (SciPy), and Systems under Test (SUT) while still offering a common interface and specification language syntax.

    In summary, PSY-TaLiRo makes the following contributions:
    \begin{enumerate}
        \item it is an open source fully modular toolbox in Python,
        \item it provides a common syntax for the temporal logic monitors, and
        \item it enables testing of Software and Hardware in the loop systems.
    \end{enumerate}
    With PSY-TaLiRo, users will be able to quickly compare different optimization and robustness computation engines without any other changes to the test setup.
    This toolbox is open-source and publicly available at:
    
    \begin{minipage}{\textwidth}
    \smallskip
    \centering
    \fbox{https://gitlab.com/sbtg/pystaliro}
    \end{minipage}

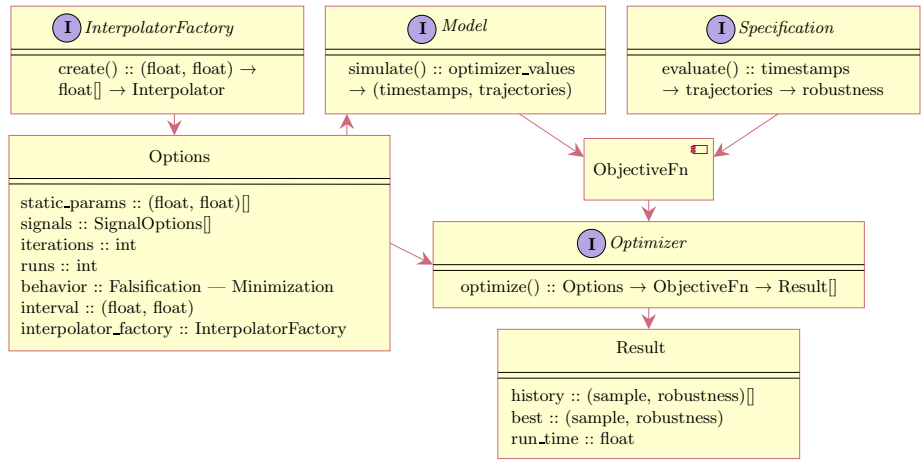
\begin{figure}[h]
    \input{arch_fig}
    \caption{Component diagram of PSY-TaLiRo architecture}
    \label{diag:architecture}
\end{figure}

\section{Architecture}
    
    The toolbox is organized into several modules: the SUT, the specification, the optimizers, and the options (see Fig. \ref{diag:architecture}). 
    Each module defines a protocol as defined in \cite{pep544} or abstract class which may be implemented or extended to create specialized implementations for a particular domain.
    The life-cycle of a test is started by providing a specification, a SUT, an optimizer and options object to the toolbox entry-point.
    Using the SUT and the specification, the toolbox generates an objective function that accepts a 1-D sequence of inputs and returns a robustness value.
    The generated objective function and the options object are then passed as parameters to the optimizer.
    The optimizer executes the objective function several times, generating and storing the input sample and the output robustness for each execution.
    When a sample is provided to the objective function, it is decomposed into a sequence of static parameters and a sequence of signals that are used as inputs to the system model.
    The output of the system model is passed to the specification, which evaluates the result and produces a robustness value, which is returned to the optimizer.
    When the optimizer is terminates its execution, a $Result$ object is returned for every execution of the optimizer in case multiple experiments are performed.

\paragraph{Type checking.}
    Optional static type checking was introduced to the Python language in version 3.6 as type annotations defined in \cite{pep484}.
    The benefit of static type checking is that multiple classes of errors can be caught before the program is executed by using a static type checker which traces the types of values through a program to ensure consistency.
    Python supports incremental typing, where a code-base can gradually add more type annotations over time instead of requiring the entire project to be typed immediately.
	PSY-TaLiRo makes extensive use of type annotations in both the internal and public APIs.
	Internally, annotations help ensure consistency between modules, reduce the difficulty of reasoning about functionality, and make it easier to implement additional features.
	For users, the annotations indicate the proper usage of the API for constructing system tests and a static type checker can provide immediate feedback.
	
	%
	%

\section{Interface}
    The PSY-TaLiRo toolbox provides a function {\tt staliro} which serves as an entry-point to the package.
    The {\tt staliro} function accepts four required parameters - a specification, a SUT, an optimizer, an options object, and one optional parameter -- an optimizer-specific options structure.
    Calling this function returns a sequence of {\tt Result} objects that store the values generated by the optimizer at each iteration and the corresponding robustness value.
    The entry-point also implements basic validation logic for its inputs, as well as the output of the SUT.

\subsection{System under Test (SUT)}
    A SUT must provide the domain-specific information required to execute or simulate a system.
    It can be a simulation model e.g., (Python, MATLAB/Simulink, etc), software-in-the-loop (SiL) (e.g., PX4, Webots, etc), or even hardware-in-the-loop (HiL).
    A SUT is responsible for accepting inputs generated by the optimizer and returning the output trajectory of the execution along with the timestamps.
    SUT may accept 3 inputs: current time,  signal values, and any initial conditions and/or static parameter values.
    Currently, PSY-TaLiRo provides two ways to run a SUT: a Blackbox class and an ODE integrator.
    The Blackbox class provides the most general way to execute a SUT because it makes no assumptions about the underlying architecture of the system it represents.
    To construct a Blackbox, a user needs to provide a function that accepts a vector of static parameters and/or initial conditions $X$, a sequence of time values $T$, and an array of signal values $U$ corresponding to each time value.
    The Blackbox function must return the time values and corresponding output/state trajectory of the SUT.
    The user also has the option to provide a set of interpolators that will translate a vector of parameters into input signals to the SUT.
    In contrast, an ODE model assumes the underlying system is represented as an ordinary differential equation and attempts to simulate the system by solving an initial-value problem.
    To construct an ODE model, a user must provide a function that accepts a time $t$, and the state at and the values of the input signal at $t$, and returns the derivatives of the system dynamics $t$.

\subsection{Specifications}
    The PSY-TaLiRo toolbox supports multiple robustness computation libraries, referred to as backends by providing a uniform interface implemented as the {\tt Specification} class.
    It is important to note that even though PSY-TaLiRo currently supports TLTk \cite{tltk} and RTAMT \cite{rtamt}, PSY-TaLiRo's modular architecure allows the user to utilize any other robustness computation engine, or, in general, any other reward or cost function.
    To construct a specification, a user must provide a system requirement written in STL, a dictionary structure specifying the requirement data, and an enumeration value specifying the back-end to use for evaluation.
    The {\tt Specification} class defines the {\tt evaluate} method, which accepts the time and signal values from the SUT and returns the robustness value.
    When the TLTk back-end is selected, the {\tt Specification} class is also responsible for parsing the STL requirement into a corresponding TLTk object representation.
    ANTLRv4 is used to generate a Python parser from a Signal Temporal Logic (STL) grammar \cite{BartocciEtAl2018survey}.
    In the case of the RTAMT back-end, no processing is done to the requirement by the {\tt Specification} class.

    Table \ref{table:1} provides an overview of the  supported common operators and syntax between the two backends.
    Beyond the common syntax, each robustness computation backend has different capabilities and the user is advised to read the respective documentation. 
    For example, TLTk supports parallel computation for scaling up to very large signals and distance based robustness \cite{FainekosP09tcs} for less conservative robustness estimates.
    On the other hand, RTAMT supports past-time operators and dense time semantics.

    
    \begin{table}[b]
        \centering
        \begin{tabular}{|l|l|}
            \hline
            \textbf{Specification Constructs} & \textbf{Syntax} \\
            \hline
            Next & \texttt{next}, \texttt{X} \\
            \hline
            Eventually & \texttt{eventually}, \texttt{F} \\
            \hline
            Globally & \texttt{always}, \texttt{G} \\
            \hline
            Until & \texttt{until}, \texttt{U} \\
            \hline
            Time constraints on operator OP & \texttt{OP[} ... \texttt{,} ... \texttt{]} \\
            \hline
            Predicates & \textit{varName} (\texttt{<=} $|$ \texttt{>=}) \textit{float} \\
            \hline
        \end{tabular}
        \caption{Common  TLTk \cite{tltk} and RTAMT \cite{rtamt} syntax supported in PSY-TaLiRo.}
        \label{table:1}
    \end{table}


\subsection{Optimizers}
    The role of the optimizer is to search for falsifying inputs to the system model by providing samples to an objective function.
    An optimizer in the PSY-TaLiRo toolbox is defined as a protocol that implements a method named optimize, which accepts an objective function, an options object, and an optional object with additional configuration options that are specific to the optimizer.
    The optimizer is also responsible for maintaining the history of samples and robustness values generated during execution and packaging them into a {\tt Result} object when completed.
    Common optimizer behavior is configured using the options object and specific optimizer behavior is configured using the optimizer-specific options object.
    PSY-TaLiRo also defines two search behaviors: falsification and minimization.
    Under falsification, the optimizer stops when the first negative robustness value is found, while minimization allows the optimizer to continue searching for lower robustness values until the execution budget is exhausted.
    The PSY-TaLiRo toolbox provides a Uniform Random Sampling optimizer and it also includes wrappers for Dual Annealing and Basinhopping optimizers.

\subsection{Options}
    To customize the behavior of the toolbox, an options object must be created and provided to the {\tt staliro} function.
    Constructing a minimally valid options object can be accomplished by providing either the {\tt static\_parameters} or {\it signals} keyword argument to the constructor. 
    The {\tt static\_parameters} attribute defines a sequence of intervals which represent the bounds of the input variables that do not change with respect to time. 
    The {\it signals} attribute represents the opposite: a sequence of signal options objects which define system inputs that vary with time.
    Other important attributes are {\tt iterations} which defines the optimizer execution budget, {\tt runs} which specifies the number of times to execute the optimizer, and {\tt interval} which specifies the interval of time for which the system should run.

\section{Examples}
    PSY-TaLiRo includes as Python demo an instance of the AircraftODE benchmark \cite{NghiemSFIGP10hscc} as well as the test setup scripts for the Python version of the F16 GCAS benchmark problem \cite{HeidlaufCBB2018f16}.
    In the following, we review how PSY-TaLiRo can interface with SUT external to Python using the Balckbox template.

\subsection{MATLAB/Simulink}
    The Simulink toolbox that is provided as a part of the MATLAB software package is useful for representing complex systems using block diagrams. 
    MATLAB additionally provides a Python library to enable access to the MATLAB engine from a Python application. 
    A PSY-TaLiRo test using a Simulink model is implemented by defining a Blackbox function which uses the MATLAB Python library to pass the parameters and signal values to the Simulink simulation engine.
    The data returned by Simulink can then be parsed into native Python data types by the Blackbox function before returning from the {\tt simulate} method.

    There are a few considerations when implementing a Blackbox that requires the MATLAB Python library. 
    Since the {\tt simulate} method of the Blackbox is called many times by the optimizer, it is very inefficient to start a new instance of the MATLAB engine every time.
    In addition, any exception that is raised during a simulation will halt the entire execution of the test, so care must be taken to ensure that any errors produced during a simulation are properly handled.

\subsection{PX4}
    The strategies used to implement a Blackbox model that can interface with the MATLAB/Simulink engine can also be applied for communication with more complex systems such as the PX4 autopilot stack \cite{PX4}.
    The PX4 is a commercial-grade autopilot software package used to control small aircraft like quad-rotors, and is capable of both SiL and HiL execution using one of several publicly available simulators.
    A successful integration of the PSY-TaLiRo toolbox and PX4 simulation environment was accomplished by using Docker to containerize the simulator and custom ground-control software to create and upload missions to the simulated drone.
    Some examples of requirements that were tested using the PX4 were to avoid exclusion zones when executing a mission, and another was to achieve a takeoff altitude within a threshold before landing.

\noindent
\begin{minipage}[t]{0.48\textwidth}
\begin{lstlisting}[frame=single, caption=Example blackbox, language=python, basicstyle=\tiny]
from staliro import models
from matlab import engine
from numpy import array

@models.blackbox()
def simulink_model(X, T, U):
    instance = engine.start_engine()
    results = instance.sim("model")
    # Timestamps
    ts = array(results[0])
    # Trajectories
    T = array(results[1])
    
    return ts, T
\end{lstlisting}
\end{minipage}
\hfill
\begin{minipage}[t]{0.48\textwidth}
\begin{lstlisting}[frame=single, caption=Example script, language=python, basicstyle=\tiny]
from tltk_mtl import Predicate
from staliro import (
    options, optimizers,
    staliro, specification
)

phi = "[]p1"
preds = {"p1": Predicate("p1", A, b)}
spec = specification.Specification(
    phi, preds)
opt = optimizers.UniformRandom()
cfg = options.StaliroOptions(
    static_parameters = [(0, 1)])
results = staliro(spec, opt, cfg)
\end{lstlisting}
\end{minipage}

\section{Conclusions}

We have presented the open-source Python toolbox PSY-TaLiRo ($\Psi$-TaLiRo).
PSY-TaLiRo implements search-based test generation for falsifying temporal logic requirements over Cyber-Physical Systems (CPS).
The toolbox is fully modular and extensible in order to accommodate different algorithms for optimization and temporal logic robustness (or arbitrary cost functions). 
Hence, PSY-TaLiRo can provide test automation support for CPS (and in particular autonomous systems) which are natively developed in Python.

\paragraph{Acknowledgements} This research was partially supported by DARPA (ARCOS FA8750-20-C-0507, AMP N6600120C4020) and NSF 1932068.

\newpage
\bibliographystyle{splncs04}
\bibliography{references}
\end{document}

%% file: arch_fig.tex
        \resizebox{\textwidth}{!}{%
            \begin{tikzpicture}[x=0.75pt,y=0.75pt,yscale=-1,xscale=1]
            
            \draw  [color={rgb, 255:red, 204; green, 106; blue, 124 }  ,draw opacity=1 ][fill={rgb, 255:red, 254; green, 254; blue, 206 }  ,fill opacity=1 ] (225,45) -- (420,45) -- (420,120) -- (225,120) -- cycle ;
            \draw    (225,75) -- (420,75) ;
            \draw    (225,79) -- (420,79) ;
            \draw  [fill={rgb, 255:red, 180; green, 167; blue, 228 }  ,fill opacity=1 ] (282.38,61.19) .. controls (282.38,55.78) and (286.78,51.38) .. (292.19,51.38) .. controls (297.61,51.38) and (302,55.78) .. (302,61.19) .. controls (302,66.61) and (297.61,71) .. (292.19,71) .. controls (286.78,71) and (282.38,66.61) .. (282.38,61.19) -- cycle ;
            \draw  [color={rgb, 255:red, 204; green, 106; blue, 124 }  ,draw opacity=1 ][fill={rgb, 255:red, 254; green, 254; blue, 206 }  ,fill opacity=1 ] (435,45) -- (638.69,45) -- (638.69,120) -- (435,120) -- cycle ;
            \draw    (435,75) -- (638.69,75) ;
            \draw    (435,79) -- (638.69,79) ;
            \draw  [color={rgb, 255:red, 204; green, 106; blue, 124 }  ,draw opacity=1 ][fill={rgb, 255:red, 254; green, 254; blue, 206 }  ,fill opacity=1 ] (300,195) -- (600,195) -- (600,255) -- (300,255) -- cycle ;
            \draw    (300,225) -- (600,225) ;
            \draw    (300,229) -- (600,229) ;
            \draw  [fill={rgb, 255:red, 180; green, 167; blue, 228 }  ,fill opacity=1 ] (400.38,210.19) .. controls (400.38,204.78) and (404.78,200.38) .. (410.19,200.38) .. controls (415.61,200.38) and (420,204.78) .. (420,210.19) .. controls (420,215.61) and (415.61,220) .. (410.19,220) .. controls (404.78,220) and (400.38,215.61) .. (400.38,210.19) -- cycle ;
            \draw  [color={rgb, 255:red, 204; green, 106; blue, 124 }  ,draw opacity=1 ][fill={rgb, 255:red, 254; green, 254; blue, 206 }  ,fill opacity=1 ] (405,137.14) -- (495,137.14) -- (495,180) -- (405,180) -- cycle ;
            \draw  [color={rgb, 255:red, 170; green, 6; blue, 58 }  ,draw opacity=1 ] (480.89,142.02) -- (490.67,142.02) -- (490.67,147.62) -- (480.89,147.62) -- cycle ;
            \draw  [color={rgb, 255:red, 170; green, 6; blue, 58 }  ,draw opacity=1 ] (479.61,142.73) -- (482.18,142.73) -- (482.18,144.12) -- (479.61,144.12) -- cycle ;
            \draw  [color={rgb, 255:red, 170; green, 6; blue, 58 }  ,draw opacity=1 ] (479.61,145.22) -- (482.18,145.22) -- (482.18,146.61) -- (479.61,146.61) -- cycle ;
            \draw [color={rgb, 255:red, 204; green, 106; blue, 124 }  ,draw opacity=1 ]   (450,255) -- (450,267) ;
            \draw [shift={(450,270)}, rotate = 270] [fill={rgb, 255:red, 204; green, 106; blue, 124 }  ,fill opacity=1 ][line width=0.08]  [draw opacity=0] (10.72,-5.15) -- (0,0) -- (10.72,5.15) -- (7.12,0) -- cycle    ;
            \draw  [color={rgb, 255:red, 204; green, 106; blue, 124 }  ,draw opacity=1 ][fill={rgb, 255:red, 254; green, 254; blue, 206 }  ,fill opacity=1 ] (345,270) -- (555,270) -- (555,360) -- (345,360) -- cycle ;
            \draw    (345,300) -- (555,300) ;
            \draw    (345,304) -- (555,304) ;
            \draw  [color={rgb, 255:red, 204; green, 106; blue, 124 }  ,draw opacity=1 ][fill={rgb, 255:red, 254; green, 254; blue, 206 }  ,fill opacity=1 ] (4.91,135) -- (270,135) -- (270,285) -- (4.91,285) -- cycle ;
            \draw    (5,165) -- (270,165) ;
            \draw    (5,168.78) -- (270,169) ;
            \draw  [fill={rgb, 255:red, 180; green, 167; blue, 228 }  ,fill opacity=1 ] (490.38,61.19) .. controls (490.38,55.78) and (494.78,51.38) .. (500.19,51.38) .. controls (505.61,51.38) and (510,55.78) .. (510,61.19) .. controls (510,66.61) and (505.61,71) .. (500.19,71) .. controls (494.78,71) and (490.38,66.61) .. (490.38,61.19) -- cycle ;
            \draw  [color={rgb, 255:red, 204; green, 106; blue, 124 }  ,draw opacity=1 ][fill={rgb, 255:red, 254; green, 254; blue, 206 }  ,fill opacity=1 ] (6.31,45) -- (210,45) -- (210,120) -- (6.31,120) -- cycle ;
            \draw    (6.31,75) -- (210,75) ;
            \draw    (6.31,79) -- (210,79) ;
            \draw  [fill={rgb, 255:red, 180; green, 167; blue, 228 }  ,fill opacity=1 ] (35.38,60.19) .. controls (35.38,54.78) and (39.78,50.38) .. (45.19,50.38) .. controls (50.61,50.38) and (55,54.78) .. (55,60.19) .. controls (55,65.61) and (50.61,70) .. (45.19,70) .. controls (39.78,70) and (35.38,65.61) .. (35.38,60.19) -- cycle ;
            \draw [color={rgb, 255:red, 204; green, 106; blue, 124 }  ,draw opacity=1 ]   (120,120) -- (120,132) ;
            \draw [shift={(120,135)}, rotate = 270] [fill={rgb, 255:red, 204; green, 106; blue, 124 }  ,fill opacity=1 ][line width=0.08]  [draw opacity=0] (10.72,-5.15) -- (0,0) -- (10.72,5.15) -- (7.12,0) -- cycle    ;
            \draw [color={rgb, 255:red, 204; green, 106; blue, 124 }  ,draw opacity=1 ]   (240,135) -- (240,123) ;
            \draw [shift={(240,120)}, rotate = 450] [fill={rgb, 255:red, 204; green, 106; blue, 124 }  ,fill opacity=1 ][line width=0.08]  [draw opacity=0] (10.72,-5.15) -- (0,0) -- (10.72,5.15) -- (7.12,0) -- cycle    ;
            \draw [color={rgb, 255:red, 204; green, 106; blue, 124 }  ,draw opacity=1 ]   (270,210) -- (297.32,223.66) ;
            \draw [shift={(300,225)}, rotate = 206.57] [fill={rgb, 255:red, 204; green, 106; blue, 124 }  ,fill opacity=1 ][line width=0.08]  [draw opacity=0] (10.72,-5.15) -- (0,0) -- (10.72,5.15) -- (7.12,0) -- cycle    ;
            \draw [color={rgb, 255:red, 204; green, 106; blue, 124 }  ,draw opacity=1 ]   (450,180) -- (450,192) ;
            \draw [shift={(450,195)}, rotate = 270] [fill={rgb, 255:red, 204; green, 106; blue, 124 }  ,fill opacity=1 ][line width=0.08]  [draw opacity=0] (10.72,-5.15) -- (0,0) -- (10.72,5.15) -- (7.12,0) -- cycle    ;
            \draw [color={rgb, 255:red, 204; green, 106; blue, 124 }  ,draw opacity=1 ]   (540,120) -- (497.5,148.34) ;
            \draw [shift={(495,150)}, rotate = 326.31] [fill={rgb, 255:red, 204; green, 106; blue, 124 }  ,fill opacity=1 ][line width=0.08]  [draw opacity=0] (10.72,-5.15) -- (0,0) -- (10.72,5.15) -- (7.12,0) -- cycle    ;
            \draw [color={rgb, 255:red, 204; green, 106; blue, 124 }  ,draw opacity=1 ]   (360,120) -- (402.5,148.34) ;
            \draw [shift={(405,150)}, rotate = 213.69] [fill={rgb, 255:red, 204; green, 106; blue, 124 }  ,fill opacity=1 ][line width=0.08]  [draw opacity=0] (10.72,-5.15) -- (0,0) -- (10.72,5.15) -- (7.12,0) -- cycle    ;
            
            \draw (304,53) node [anchor=north west][inner sep=0.75pt]   [align=left] {\textit{Model}};
            \draw (289,55) node [anchor=north west][inner sep=0.75pt]   [align=left] {\textbf{I}};
            \draw (240,82) node [anchor=north west][inner sep=0.75pt]   [align=left] {{\small simulate() :: optimizer\_values }\\{\small → (timestamps, trajectories)}};
            \draw (511,54) node [anchor=north west][inner sep=0.75pt]   [align=left] {\textit{Specification}};
            \draw (458,82) node [anchor=north west][inner sep=0.75pt]   [align=left] {{\small evaluate() :: timestamps }\\{\small → trajectories → robustness}};
            \draw (421,204) node [anchor=north west][inner sep=0.75pt]   [align=left] {\textit{Optimizer}};
            \draw (406.5,205) node [anchor=north west][inner sep=0.75pt]   [align=left] {\textbf{I}};
            \draw (318,234) node [anchor=north west][inner sep=0.75pt]   [align=left] {{\small optimize() :: Options → ObjectiveFn → Result[]}};
            \draw (410,153) node [anchor=north west][inner sep=0.75pt]   [align=left] {ObjectiveFn};
            \draw (426,277) node [anchor=north west][inner sep=0.75pt]   [align=left] {Result};
            \draw (353,310) node [anchor=north west][inner sep=0.75pt]   [align=left] {{\small history :: (sample, robustness)[]}\\{\small best :: (sample, robustness)}\\{\small run\_time :: float}};
            \draw (101,144) node [anchor=north west][inner sep=0.75pt]   [align=left] {Options};
            \draw (12,175) node [anchor=north west][inner sep=0.75pt]   [align=left] {{\small static\_params :: (float, float)[]}\\{\small signals :: SignalOptions[]}\\{\small iterations :: int}\\{\small runs :: int}\\{\small behavior :: Falsification | Minimization}\\{\small interval :: (float, float)}\\{\small interpolator\_factory :: InterpolatorFactory}};
            \draw (497,55) node [anchor=north west][inner sep=0.75pt]   [align=left] {\textbf{I}};
            \draw (56,54) node [anchor=north west][inner sep=0.75pt]   [align=left] {\textit{InterpolatorFactory}};
            \draw (38,82) node [anchor=north west][inner sep=0.75pt]   [align=left] {{\small create() :: (float, float) → }\\{\small float[] → Interpolator}};
            \draw (42,54) node [anchor=north west][inner sep=0.75pt]   [align=left] {\textbf{I}};

            \end{tikzpicture}

        }